\documentclass[prd,aps,floats,twocolumn,nofootinbib]{revtex4-1}
\usepackage{slashed}
\usepackage{mathtools}
\usepackage{amsfonts}
\usepackage{amssymb}
\usepackage{epsfig}
\usepackage{rotating}
\usepackage{url}
\usepackage{times}
\usepackage{color}
\usepackage{bm}
\usepackage{xcolor,colortbl}
\usepackage{hyperref}
\usepackage[alsoload=hep]{siunitx}
\usepackage{enumitem}
\usepackage{fancyhdr}
\usepackage{anyfontsize}
\usepackage{wasysym}
\usepackage{empheq}

\begin{document}

\newcommand{\nn}{\nonumber}
\newcommand{\ph}{\phantom}
\newcommand{\eps}{\epsilon}
\newcommand{\be}{\begin{equation}}
\newcommand{\ee}{\end{equation}}
\newcommand{\bea}{\begin{eqnarray}}
\newcommand{\eea}{\end{eqnarray}}

%%%%%%%%%%%%%%%%%%%%%%%%%%%%%

\title{Parity violating Friedmann Universes}

\author{Jo\~{a}o Magueijo}
\email{j.magueijo@imperial.ac.uk}
\affiliation{Theoretical Physics Group, The Blackett Laboratory, Imperial College, Prince Consort Rd., London, SW7 2BZ, United Kingdom}
\author{Tom Z\l o\'{s}nik}
\email{zlosnik@fzu.cz}
\affiliation{CEICO, Institute of Physics of the Czech Academy of Sciences, Na Slovance 1999/2, 182 21, Prague}

\date{\today}

\begin{abstract}
We revisit extensions of the Einstein-Cartan theory where the cosmological constant $\Lambda$ is promoted to a variable, at the cost of allowing for torsion even in the absence of spinors. We remark that some standard notions about FRW Universes collapse in these theories, most notably spatial homogeneity and isotropy may now co-exist with violations of parity invariance. The parity violating solutions have non-vanishing Weyl curvature even within FRW models.
The presence of parity violating torsion opens up the space of possible such theories with relevant FRW modifications:
in particular the Pontryagin term can play an important role even in the absence of spinorial matter. 
We present a number of parity violating solutions with and without matter. The former are the non-self dual vacuum solutions long suspected to exist. The latter lead to tracking and non-tracking solutions with a number of observational problems, unless we invoke the Pontryagin term. 
An examination of the Hamiltonian structure of the theory reveals that the parity even and the parity violating solutions belong to two distinct branches of the theory, with different gauge symmetries (constraints) and different numbers of degrees of freedom. 
The parity even branch is nothing but standard relativity with a cosmological constant which has become pure gauge under conformal invariance if matter is absent, or a slave of matter (and so not an independent degree of freedom) if non-conformally invariant 
matter is present.  
%New effects, nonetheless, exist if we project all the results onto the (non-conformally invariant) matter frame. 
In contrast, 
the parity violating branch contains a genuinely new degree of freedom. 
\end{abstract}

\maketitle

\section{Introduction}

In previous work~\cite{alex1,alex2} 
the possible variability of the cosmological ``constant'', $\Lambda$, was examined from two standpoints. 
First, there is the apparent obstruction presented by Bianchi identities to the variability of the $\Lambda$ term in the
Einstein equations, and the fact that this obstacle is promptly removed in the first order formalism if torsion is permitted. 
That the presence of torsion can change dramatically the perspective of problems is not new (see for example~\cite{Dolan:2009ni,Mercuri:2009zt,Baekler:2010fr,Poplawski:2011j,Schucker:2011tc,Cid:2017wtf,Zlosnik:2018qvg,Flanagan:2003rb,Sotiriou:2005hu,Bauer:2010jg,Farnsworth:2017wzr,Addazi:2017qus}). In the context of~\cite{alex1}
the requisite torsion is provided by a topological term (the Euler invariant) multiplied by an appropriate function of $\Lambda$
uniquely specified by the Bianchi identities. Then, not only is a varying $\Lambda$ allowed, but in the absence of matter and Weyl curvature, $\Lambda$ is left unspecified by the field equations and is totally free. 

More importantly, there was a second perspective, put forward in~\cite{alex0} and~\cite{LeeJ}. Could $\Lambda$, or a function thereof, be promoted to a dynamical variable canonically conjugate to the imaginary part of the Chern-Simons invariant? The latter is a well-known measure of time capable of surviving quantum gravity~\cite{chopin}. The prospect of a quantum complementarity principle between cosmological time and $\Lambda$ leads to interesting speculations regarding the possible disruption to an omnipresent time-line for our Universe~\cite{LeeJ,Gott:1997pm}.  More mundanely it is natural to ask: can such quantum complementarity arise from a classical Poisson bracket
in a well defined Hamiltonian theory? 

Regrettably the minimal realization explored in~\cite{alex1,alex2} leads to a drastically unviable cosmology. This is not altogether surprising. Against current trends, the proposed new theory of gravity has {\it fewer} free parameters than General Relativity (GR), instead of 
a multitude of new knobs that can be turned at will to fit any dataset, as is often the case with modifications to GR. No wonder a preliminary investigation ends in phenomenological disaster~\cite{alex2}. 

In this paper we revisit the work of~\cite{alex1,alex2}  and push it beyond a first exploration. As a central result, 
we show that torsion
opens up the doors to non-parity invariant homogeneous and isotropic cosmological models. These live on a separate 
branch from the parity even solutions reported in~\cite{alex2}. The two branches have qualitatively different equations
and this is reflected in the different structures of Hamiltonian constraints, and even different number of degrees of freedom.
It can be argued that they constitute separate theories; or, at the very least,
are two entirely independent phases of the same theory. 

It is not difficult to see that such solutions may exist, and that parity invariance is not a symmetry necessity for homogeneous and isotropic Universes, but, rather, results from the field equations that impose zero torsion in standard GR.
Following the notation of~\cite{alex2}, homogeneity, isotropy (and the vanishing of spatial curvature) imply the tetrad:
\bea
e^0&=&dt\label{e0}\\
e^i&=&adx^i\label{ei}
\eea
where $a(t)$ is the expansion factor, $t$ is proper cosmological time and $x^i$ are comoving cartesian coordinates (later in this paper we shall reinstate spatial 
curvature). No violation of parity is allowed for the tetrad. However, 
the torsion $T^a=De^a \equiv de^{a}+\omega^{a}_{\ph{a}b}e^{b}$ is a 2-form, so using only homogeneity and isotropy it can take the general form:
\bea
T^0&=&0\label{T0}\\
T^i&=&-T(t)e^0 e^i+P(t)\epsilon^{i}_{\ph{i}jk} e^j e^k.\label{Ti}
\eea
The term in $P(t)$ is parity-odd and can be excluded if we impose parity invariance in addition to homogeneity and isotropy
(as in~\cite{alex2}), but not
otherwise. This term was first introduced by Cartan~\cite{cartan} and has been considered more recently in~\cite{Baekler:2010fr,spiral}.
The new solutions in our paper result directly from this term. 
They imply terms in the curvature that drop out of Einstein's equations. The parity-odd torsion $P$ produces Weyl as well as Ricci curvature in FRW, whereas $T$ only produces Ricci curvature. Thus, the vanishing of the  Weyl tensor in homogeneous and isotropic Universes is also not a symmetry requirement, but an implication of the field equations in certain theories (but not in~\cite{alex1}). 

As we shall see below, the parity-odd term in the torsion not only reveals a new branch in the Hamiltonian structure of the theory 
proposed in~\cite{alex1,alex2}, but opens up the space of possible theories with relevance for homogeneous and 
isotropic models. An action  term proportional to the Pontryagin invariant was discounted in~\cite{alex2} because it vanishes identically
for parity-even FRW models. The possibility of parity violating solutions brings this term into play. In Section~\ref{SDtheories} we
propose a possible construction where the Immirzi parameter $\gamma$ controls the strength of the Pontryagin term. 
In Section~\ref{EOMFRW} we derive the equations of motion and reduce them to FRW Universes. 
We find that indeed new terms appear in the FRW equations when $\gamma$ is finite, due to the Pontryagin and the $P$ terms.
In view of the complexity of the equations we defer to future work the detailed study of these Pontryagin homogeneous and isotropic cosmologies.
The rest of the paper assumes $|\gamma|\rightarrow \infty$ and is organized as follows. 

In Section~\ref{newsols} we present solutions to the parity violating branch of the FRW models studied in~\cite{alex1,alex2}. For vacuum solutions we find the non-self dual solutions previously suspected to exist. In the presence of matter we find the tracking and non-tracking solutions, presenting also some preliminary simple results for finite $\gamma$. We identify several observational 
problems and the way to fix them. We conclude that finite $\gamma$ is needed for a viable cosmology.

In the remaining Sections~\ref{Hamiltonian}, \ref{confsym} and~\ref{secondclass} we find the Hamiltonian structure of the theory. We show that setting $P=0$ or not leads to different theories, with different numbers of constraints and degrees of freedom. We identify the presence of an extra constraint with respect to GR if $P=0$, and prove that it represents conformal invariance. Switching on the parity breaking term in 
the torsion amounts to breaking conformal invariance. The solutions found in~\cite{alex2}, therefore, are nothing but conformal gauge transformations performed upon GR, unless non-conformal matter is added. In contrast our new solutions represent a genuinely new degree of freedom. 

%In a concluding Section we summarize our results and present some final insights. 

\section{Theories with a varying-$\Lambda$ and self-duality}\label{SDtheories}
In the exploration of theories with a variable $\Lambda$ in~\cite{alex0,alex1}, it 
was found that the duality interchanging
\be\label{dual}
R^{ab} \leftrightarrow \frac{\Lambda(x)}{3}e^a e^b,
\ee
plays a central role. This duality  is satisfied by the solutions to
the Einstein equation in the absence of matter and Weyl curvature:
\be\label{SDequation}
R^{ab}=\frac{\Lambda}{3}e^a e^b
\ee
and these are called self-dual (SD) solutions. It was noted in \cite{alex1}
that the SD condition  trivially survives the promotion of $\Lambda$ to a field. 
One may ask what is the most general gravitational action which
\begin{itemize}
\item[1)] is at most quadratic in the curvature; 
\item[2)] leaves the standard Einstein equations (the $e$ equation) unmodified; 
\item[3)] contains the Palatini action as a term (or reduces to standard torsion-free GR should $\Lambda$ be constant); 
\item[4)] is SD, remains invariant under duality (\ref{dual})?
\end{itemize}
A possible answer is:
\begin{widetext}
\be\label{SDcond}
S^g=-\int \frac{3}{2\Lambda}\left( \epsilon_{abcd}+\frac{2}{\gamma}\eta_{ac}\eta_{bd}\right)\left(R^{ab}-\frac{\Lambda}{3}e^a e^b\right)
\left(R^{cd}-\frac{\Lambda}{3}e^c e^d\right)-\frac{2}{\gamma}\int T^aT_a.
\ee
\end{widetext}
This action is evidently symmetric under (\ref{dual}). 
It has the property that it satisfies a stronger version of 3), in that it reduces to the Holst action for constant $\Lambda$. 
It can be unwrapped into four terms $S^g=S_{Pal}+S_{Eul}+S_{NY}+ S_{Pont}$ according to:
\bea
S_{Pal}&=&\int  \epsilon_{abcd}\left( e^a e^b R^{cd} -\frac{\Lambda}{6} e^a e^be^ce^d\right)  \label{palatini},\\
S_{Eul}&=&-\frac{3}{2}\int \frac{1}{\Lambda} \epsilon_{abcd}R^{ab}R^{cd} ,\\
S_{NY}&=&\frac{2}{\gamma}\int e^ae^b R_{ab}-T^aT_a ,\\
S_{Pont}&=&-\frac{3}{\gamma}\int \frac{1}{\Lambda} R^{ab}R_{ab} .
\eea
The first term is the usual Palatini action, as required by point 3. The second term is the 
quasi-Euler term in the form contemplated in~\cite{alex1}. The pre-factor $-\frac{3}{2\Lambda}$ is fully determined by the duality.
Following the ``quasi-topological principle'' advocated in~\cite{alex1} (see Section II of that reference) 
one could add other quasi-topological terms preserving the duality. The term
$S_{NY}$ is the Nieh-Yan invariant. Its first term is associate the Holst term, with $\gamma$ the Immirzi parameter. To this term one must generally add the ``torsion-squared'' second term, to ensure it forms a boundary term (and comply with our requirement 2). 
If this term is present, then 
the simple form of the action Eq.~(\ref{SDcond}) (which is just a special realization of the SD condition, albeit one with particular aesthetic appeal)
requires the presence of the term $S_{Pont}$. This is the Pontryagin invariant multiplied by $-\frac{3}{\gamma \Lambda}$.
The last two terms vanish if the Immirzi parameter $\gamma$ is infinite \footnote{The theory in the limit of infinite $\gamma$ also corresponds to a specific gauge theory of gravity based on de Sitter or anti de Sitter groups in a symmetry broken phase with remnant Lorentz symmetry \cite{Mignemi:1998us,Magueijo:2013yya,Westman:2014yca}.}. If $\gamma$ is finite then the action (\ref{SDcond}) does not transform homogeneously under a parity transformation acting on fields with local Lorentz indices.

We stress that this pleasant looking form of the action is not the most general one. The first term in (\ref{SDcond}) 
is manifestly SD and built from basic blocks ($\eta_{ab}$ and $\epsilon_{abcd}$), but to make it comply with condition 3
we had to subtract the torsion squared second term in  (\ref{SDcond}). But we could equally well have subtracted the 
the Holst term contained in the first term in  (\ref{SDcond}), so that the proposed action would just differ from that proposed in~\cite{alex1,alex2} by the Pontryagin invariant multiplied by a function of
$\Lambda$. Once the connection with the Holst term is lost, the factor multiplying the Pontryagin could be any function of $\Lambda$. 
This results from the fact that the 
Pontryagin term already is invariant under (\ref{dual}), or rather, the term needed to make it manifestly invariant is 
zero (it is proportional to $e^a e^b e_ae_b$).  The pre-factor could be any constant divided by $\Lambda$ on dimensional grounds. 
If the constant $\kappa=8\pi G$  can be used here the pre-factor could also be any power of $\Lambda$.  In spite of this, for simplicity
we shall use (\ref{SDcond}) for the rest of this paper. 

Eq.~(\ref{SDcond}) is the proposed gravity action. Matter can be added to it as usual:
\be
S=\frac{1}{2\kappa}S^g(e,\omega,\Lambda)+S^d(\Phi, e),
\ee
where we assume that the matter Lagrangian does not depend on $\omega$. 
The stress-energy 3-form is given by:
\be
\tau_a =\frac{1}{2}\frac{\delta S_M}{\delta e^a}
\ee
and no spin current is present generating torsion from matter.  This is likely to be a very good approximation in cosmology and is exact if spinors are assumed to couple to the torsion-free spin-connection.

%It was found in~\cite{alex1} that the theory comprised by $S=S_{Pal}+S_{Eul}$ is enough to render a varying Lambda consistent with the condition (\ref{SD}).  Indeed for solutions to (\ref{SD}) the terms $S_{Eul}$ generates just the right torsion that can be inferred from Bianchi identities applied to (\ref{SD}) itself:
%\be
%T^a=-\frac{1}{2\Lambda}d\Lambda e^a.
%\ee
%Furthermore, it can be checked that $S_{NY}$ and $S_{Pont}$ vanish identically for SD solutions. Addition of matter in the context of  a homogeneous, isotropic and parity even background was explored in~\cite{alex2}. It can also be checked that  $S_{NY}$ and $S_{Pont}$  vanish for these space-times. 

%, they must vanish in any FRW reduction in which one imposes parity invariance as well as homogeneity and isotropy. The point of this paper is that such extra condition is unwarranted if we consider theories with torsion, as is the case here. Then there will be solutions with parity-even and parity-odd torsion. The latter will bring the new terms into the problem. They will allow for non-SD solutions for pure Lambda (no matter). In the presence of matter, they convert the equation in Lambda from an algebraic equation into a ODE. [RESULTS.... New scaling solutions will be possible. The no-go theorem for a radiation epoch in this theory is bypassed (ah, ah, wishful thinking)] 

%FRW with torsion: results usually quoted are not the result of symmetries, but of equations of motion peculiar to Einstein gravity. If torsion is allowed they are not true. This is specifically true of 

\section{The field equations and their FRW reduction}\label{EOMFRW}
It is easy to verify that variation of the action (\ref{SDcond}) with respect to $e$, $\omega$ and $\Lambda$ leads to:
\bea
\epsilon_{abcd}{\left(e^b  R^{cd}-\frac{1}{3}\Lambda e^b  e^c  e^d\right)}=-2 \kappa \tau_a\label{Eeq}\\
T^{[a}  e^{b]}=-\frac{3}{2\Lambda^2}d\Lambda  R^{ab}+\frac{3}{4\gamma \Lambda^2}\epsilon^{abcd}d\Lambda R_{cd}\label{omEq}\\
\epsilon_{abcd}{\left(R^{ab}  R^{cd}-\frac{1}{9}\Lambda^2 e^a   e^b  e^c  e^d\right)}+\frac{2}{\gamma}R^{ab}R_{ab}=0\label{Leq}
\eea
Since the Nieh-Yan term is a topological term (rather than a quasi-topological term) it does not contribute to
any of the three field equations. As promised by the need to satisfy requirement 3, the first equation is the unmodified Einstein equation. 
Thus, only the last 2 equations receive new terms with respect to the equations in~\cite{alex1,alex2}, and these arise solely from the quasi-topological Pontryagin term $S_{Pont}$. For action (\ref{SDcond}) they are tied to the Immirzi parameter (but see the proviso explained in Section~\ref{SDtheories}), and vanish for $\gamma\rightarrow \infty$.

We can now reduce these equations to a homogeneous and isotropic Universe using (\ref{e0}) and (\ref{ei}) for the tetrad $e^a$,
and (\ref{T0}) and (\ref{Ti}) for the torsion $T^a$. 
Then, the definition $T^a\equiv D e^a=de^a+\omega^a_{\;b} e^b$ 
implies: 
\bea
\omega^i_{\; 0}&=&g(t) e^i =\left(\frac{\dot a}{a}+T\right)e^i\label{omega0i}\\
\omega^{ij}&=&-P\epsilon^{ijk}e^k,\label{omegaij}
\eea
where for the time being we have assumed vanishing spatial curvature and 
with the usual Hubble parameter replaced by the function:
\be\label{gfunc}
g=\frac{\dot a}{a}+T,
\ee
- where dots denote derivatives with respect to cosmic proper time - and the parity-odd component of the torsion, $P$, inducing the spatial components $\omega^{ij}$ of the connection. 
Therefore, the curvature 2-form $R^{ab}\equiv d\omega^{ab}+\omega^a_{\;c}\omega^{cb}$
has  components:
\bea
R^{0i}&=&\frac{(ag)^.}{a}e^0  e^i + gP\epsilon^{ijk}e^je^k \label{F0i}\\
R^{ij}&=&(g^2-P^2)  e^i  e^j -\frac{(aP)^.}{a}\epsilon^{ijk}e^0 e^k\label{Fij}
\eea
and we see that no longer do $R^{0i}\propto e^0 e^i$ and $R^{ij}\propto e^i e^j$, with possibly different 
proportionality constants, as is the case for parity invariant solutions (with implications studied elsewhere). 
By looking at the projection $\epsilon_{abcd} e^b R^{cd}$
we can find the Ricci and Weyl components of the curvature, ${\cal R}^{ab}$ and ${\cal W}^{ab}$, respectively. 
With $R^{ab}={\cal R}^{ab}+{\cal W}^{ab}$ we have: 
\bea
{\cal R}^{0i}&=&\frac{(ag)^.}{a}e^0  e^i \\
{\cal R}^{ij}&=&(g^2-P^2)  e^i  e^j
\eea
for the Ricci component entering the Einstein equation, and
\bea
{\cal W}^{0i}&=&gP\epsilon^{i}_{\ph{i}jk}e^je^k \label{Weyl1}\\
{\cal W}^{ij}&=& -\frac{(aP)^.}{a}\epsilon^{ij}_{\ph{ij}k}e^0 e^k\label{Weyl2}
\eea
for the Weyl curvature. 
Whereas the parity-even part of the torsion, $T$, contributes only to Ricci curvature, 
the parity-odd component, $P$, contributes to both Ricci and Weyl curvature.  It is therefore possible to have
Weyl curvature in FRW models, if the appropriate form of (parity-odd) torsion is present.
It is the presence of Weyl in FRW models that allows for the novelties found in this paper.

Avoiding simplifications that would obscure comparison with previous work, 
the full set of equations (\ref{Eeq})-(\ref{Leq})  applied to FRW becomes:
\begin{widetext}
\bea
g^{2}-  P^{2} &=&\frac{\Lambda+\kappa\rho }{3}\label{EeqF}\\
\frac{(ag)^.}{a}&=& \frac{\Lambda}{3}-\frac{\kappa}{6}(\rho +3p)\label{ray}\\
T&=&\frac{\dot\Lambda}{2\Lambda^2}\left(\Lambda+ \kappa\rho -\frac{6}{\gamma}gP\right)\label{TeqF}\\
P&=&\frac{3\dot{\Lambda}}{\Lambda^2}\left(gP +\frac{\Lambda +\kappa\rho}{6\gamma}\right)\label{PeqF}\\
(\Lambda+\kappa\rho)
\left(\Lambda-\frac{\kappa}{2}(\rho +3p)\right)-\Lambda^2&=&18g
 P\frac{(a P)^.}{a} 
+\frac{9}{\gamma}\left(\frac{\Lambda+\kappa\rho}{3}\frac{(aP)^.}{a} +\frac{2}{3}\left(\Lambda-\kappa\frac{\rho+3p}{2}\right)gP\right)
\label{LeqF}
\eea
\end{widetext}
It can be checked that differentiation of the equation (\ref{EeqF}) and the use of equations (\ref{ray})-(\ref{LeqF}) implies that 

\begin{align}
\dot\rho+3\frac{\dot a}{a}(\rho+p)=0. \label{cons}
\end{align}
and this equation can be used to replace (\ref{ray}) to produce a very practical complete set of equations. As already noted in~\cite{alex2}, Eq.~\ref{cons} implies that matter is covariantly conserved with regards to the torsion-free connection. 
As explained in~\cite{alex2}, this result is actually very general, and can be proved using Noether's theorem and an appropriate set of assumptions. 

By letting $\gamma\rightarrow\infty$ we see that we can set $P=0$ and recover the equations (31)-(34) of reference~\cite{alex2}. However, the parity even solutions with $P=0$ are merely a branch of the theory, and we can explore more general solutions
with $P\neq 0$  and investigate how these relate to the original ones found in~\cite{alex2}. Furthermore if $\gamma$ is finite we cannot 
consistently set $P=0$ (except for trivial solutions, which are briefly discussed in Section \ref{finitegamham}) if is finite. Thus the presence of the Pontryagin term in the action precludes the existence of parity invariant solutions. 

Even before we start looking in the detail at the new solutions, we can see that, with the introduction of $P$, the system of equations obtained is qualitatively very different. Crucially, no longer is the Lambda equation (\ref{LeqF}) an algebraic equation, a fact behind much of the behaviour reported in~\cite{alex2}. This is directly due to the presence of Weyl tensor in parity violating solutions. 
Without Weyl curvature, the Euler term in (\ref{Leq}) (first term) can be eliminated in terms of $\Lambda$ and $\rho$ and $p$ via the Einstein equation, and the last   (Pontyagrin)  term in  (\ref{Leq}) vanishes. If Weyl curvature is present, it contributes to the Euler invariant with terms that  cannot be determined by the Einstein equations in terms of matter and $\Lambda$ (recall the Enstein equation remains unmodifed and so dependent only on the Ricci tensor). Also the Pontryagin term no longer vanishes (and neither can it be eliminated by the Einstein equation). The result is a differential equation,  as evidenced by 
the right hand side of (\ref{LeqF}). The Lambda equation is only algebraic if Weyl curvature vanishes, as is the case with parity
even solutions.

%We examine the novelties arising by the parity violating solutions below, separating the finite and infinite $\gamma$ cases.   

\section{The new branch of the $\gamma\rightarrow\infty$ theory}\label{newsols}
In this paper we focus on parity breaking solutions 
without Pontryagin term. 
By letting $\gamma\rightarrow\infty$ equations (\ref{EeqF})-(\ref{LeqF}) simplify to:
\bea
g^{2}-  P^{2} &=&\frac{\Lambda+\kappa\rho }{3}\\
\frac{(ag)^.}{a}&=& \frac{\Lambda}{3}-\frac{\kappa}{6}(\rho +3p)\\
T&=&\frac{\dot\Lambda}{2\Lambda}\left(1+\frac{\kappa\rho}{\Lambda}\right)\\
P&=&\frac{3\dot{\Lambda}}{\Lambda^2}gP\label{Peq}\\
18g
 P\frac{(a P)^.}{a}
&=&
(\Lambda+\kappa\rho)
\left(\Lambda-\frac{\kappa}{2}(\rho +3p)\right)-\Lambda^2.\eea
Setting $P=0$ we recover equations 31-34 in~\cite{alex2}, as already stated. Specifically, we see that without matter we have
$\Lambda^2=\Lambda^2$ for the equation of motion (EOM) of $\Lambda$, signalling that it remains unspecified. Whether this is a true degree of freedom 
will be examined in Sections~\ref{Hamiltonian} and~{\ref{confsym}. If there is matter, then the Lambda algebraic EOM implies that
$\Lambda$ tracks matter, except in the case of pure radiation, where the equation sets the radiation density to zero,
and we are back to pure $\Lambda$. Again, we shall re-examine this in the next Section, and understand it in relation to the
conformal invariance of radiation. Finally, we note that Eq.~(\ref{Peq}) is absent for $P=0$, since it reduces to $0=0$.

However, if $P\neq 0$ an entirely new branch of the system emerges, with qualitatively different behaviour. 
We gain a new differential equation, since (\ref{Peq}), rather than collapsing to $0=0$, now states:
\be\label{PeqF1}
1=\frac{3\dot{\Lambda}}{\Lambda^2}g. 
\ee
In addition the equation obtained by varying $\Lambda$ is no longer algebraic, as announced. The rest of this Section is devoted 
to finding the new solutions, and the next Section to examine the meaning and orgin of the new branch. 

In Appendix I we present an alternative derivation of the field equations following from the action (\ref{SDcond}). 
One starts by writing down the action for FRW and only then variations are taken. This will be useful for an examination of the 
Hamiltonian structure later. A first order closed system can then be found. 

\subsection{Solutions without matter}
\label{solutions_without_matter}
We now consider solutions where $P\neq 0$ (parity violation), but $\Lambda\neq 0$ and $\rho=p=0$. These are the non-SD vacuum solutions speculated to exist  in ~\cite{alex1,alex2}. From now on, for convenience,  we will often use ``conformal'' versions of both $P$ and $g$ defined from:
\bea
c&=&Pa\\ 
b&=& ga.
\eea
It can be shown (see Appendix I) that the equations of motion then reduce to
\bea
\frac{da}{dt}&=& b- \frac{a^{2}\Lambda}{6b}\label{aeq}\\
\frac{db}{dt} &=&  \frac{1}{3}a \Lambda\label{beq} \\
\frac{d\Lambda}{dt} &=& \frac{a\Lambda^{2}}{3b}\label{Leq}\\
\frac{dc}{dt}& =& 0\label{c-eq}
\eea
with Hamiltonian constraint:
\begin{align}
k-c^{2}+b^{2} =  \frac{1}{3}a^{2}\Lambda
\end{align}
where we have allowed for a non-zero spatial curvature $k$. Eq.~(\ref{c-eq}) implies constancy of the conformal version of $P$:
\be
c=c_{0},
\ee 
so we see that the parity violating term behaves like negative curvature in the Hamiltonian constraint. 
Also, Eqs.~(\ref{beq}) and (\ref{Leq}) can be combined to yield 
\be
b = b_{0}\Lambda .
\ee 
Then, the $\Lambda$ evolution equation combined with the Hamiltonian constraint can be used to obtain an ordinary differential equation for $\Lambda$ that can be integrated to yield, for $k-c_{0}^{2}>0$ and $b_{0}>0$, the parametric expression:
\begin{align}
\Lambda &= \frac{\sqrt{k-c_{0}^{2}}}{b_{0}} \sinh\bigg(\eta-\eta_{0}\bigg) \label{l1} \\
a\sqrt{\Lambda/3} &= \sqrt{k-c_{0}^{2}}\cosh\bigg(\eta-\eta_{0}\bigg)
\end{align}
where $\eta_{0}$ is an integration constant and we've define the coordinate $\eta$ via $d\eta = \sqrt{\Lambda/3}dt$ (not conformal time, therefore). For the case $c_{0}^{2}-k>0$ and $b_{0}>0$ there is instead the implicit solution

\begin{align}
\eta-\eta_{0} &= \ln \bigg|\sqrt{\frac{b_{0}^{2}\Lambda^{2}}{c_{0}^{2}-k}-1}+\frac{b_{0}\Lambda}{\sqrt{c_{0}^{2}-k}}\bigg| \label{l2}
\end{align}
Thus for $\eta\gg \eta_{0}$ in both cases, $\Lambda$ increases as $e^{\eta}$. 

These are the non-SD vacuum solutions of the theory when $\gamma\rightarrow \infty$. We notice that indeed they 
have non-vanishing Weyl tensor, 
even though they are homogeneous and isotropic (but not parity invariant). 
From (\ref{Weyl1}) and (\ref{Weyl2}) we see that ${\cal W}^{ij}=0$ for these solutions, but ${\cal W}^{01}\neq 0$. 
Because the Weyl tensor is non-zero, Einstein's equations in vacuum no longer reduce to Eq.~(\ref{SDequation}).

We defer to future work a complete study of the equivalent solutions under the influence of the Pontryagin term. 
As a brief illustration of differences that can arise when $\gamma$ is finite, we note that for this case the following solution exists \footnote{This is a solution that corresponds to a foliation of de Sitter space with $k=0$; the general solution for $a$ involves a decaying exponential that allows for the correct $k >0$ and $k<0$ forms.}:

\begin{align}
\Lambda &= \Lambda_{0}\\
c &=0 \\
a &= e^{\sqrt{\frac{\Lambda_{0}}{3}}t}
\end{align}
i.e. this solution corresponds to de Sitter space and represents as a General Relativistic limit as $\Lambda$ is constant and the torsion vanishes. Now consider small homogeneous perturbations around this solution, for example $\Lambda(t) = \Lambda_{0}+  \lambda(t)$, $c(t) = \delta c(t)$. It can then be shown that

\begin{align}
\delta c (t) &= e^{-2\sqrt{\frac{\Lambda_{0}}{3}}t}{\cal C}_{1}\\
\lambda(t) &= -\frac{2\gamma \sqrt{\Lambda_{0}}}{\sqrt{3}} e^{-\sqrt{3\Lambda_{0}}t}{\cal C}_{1} + {\cal A}_{1} 
\end{align}
Therefore, this solution is stable against small time-dependent perturbations.

%END TGZ %

\subsection{Tracking solutions in the presence of matter}\label{tracksolutions}
It was found in~\cite{alex1} that when $c=0$, matter dramatically altered the behaviour of the theory. We now investigate the effect of matter when $Pa = c\neq 0$. We provisionally set $k=0$ (in effect assuming that curvature provides a negligible contribution to the Einstein equations) and look for solutions where the gravitational fields in some sense scale in the same way as the dominant matter component. We take this component to be a perfect fluid with pressure $p=w\rho$. From the matter conservation equation, Eq.~\ref{cons}), it follows that $\rho \sim a^{-3(1+w)}$. We adopt a notation:
\begin{align}
\frac{8\pi G \rho}{3}  &\equiv \Omega_{w} a^{-3(1+w)}
\end{align}
and we make the ``scaling''  or ``tracking'' ansatz:
\begin{align}
\Lambda &=  3\Omega_{\Lambda} a^{-3(1+w)} \\
c^{2} = (Pa)^{2} &= \Omega_{c}a^{-(1+3w)} \label{cansatz} \\
b^{2} = (ga)^{2} &= \Omega_{b}a^{-(1+3w)} 
\end{align}
where the $\Omega_{i}$ are constants. The Hamiltonian constraint equation then reads:
\begin{align}
\Omega_{b} &= \Omega_{w} + \Omega_{\Lambda}+ \Omega_{c}
\end{align}
In the limit $\gamma\rightarrow \infty$, the system of equations of motion (see Appendix I) becomes a set of algebraic equations 
for the $\Omega_i$. The solutions can be categorized as follows, with the subscripts enumerating various cases:
\begin{align}
\Omega_{(1)\Lambda} &= \frac{3(1+w)(1+3w)}{5+3w}\Omega_{w} ,\quad \Omega_{(1)c} = -\frac{1}{2}(2+3w)\Omega_{w}
\end{align}
\begin{align}
\Omega_{(2)\Lambda} &=0 ,\quad \Omega_{(2)c} = -\Omega_{w} \\
& \nn\\
\Omega_{(3)\Lambda} &=0 ,\quad 
\Omega_{(3)c} = \frac{1}{2}\Omega_{w} \label{infinitetrack}
\end{align}
\begin{align}
\Omega_{(4)\Lambda} &= -\frac{3(1+w)}{1+3w}\Omega_{w}, \quad \Omega_{(4)c} = \frac{2}{1+3w}\Omega_{w}
\end{align}
For Case 1, $\Omega_{c}$ is negative for $w>-2/3$ and hence - from (\ref{cansatz}) - $c \in i\mathbb{R}$ and therefore the solution is unphysical for cosmologies possessing periods of matter or dust domination. Similarly Case 2 is excluded and whilst for Cases 3 and 4, $\Omega_{c}$ is positive, its magnitude is too great to give a realistic expansion history of the Universe.

Thus, the scaling solution for infinite $\gamma$ are as unviable as those studied in~\cite{alex2}.
We defer to future work a more complete study of the finite $\gamma$ case, but here give a preliminary result. 
It may be shown that in the case of finite $\gamma$, the quantities $R_{\Lambda} \equiv \Omega_{\Lambda}/\Omega_{w}$, $R_{c}\equiv \Omega_{c}/\Omega_{w}$ obey the equations:

\begin{widetext}
\begin{align}
R_{\Lambda}\bigg(\frac{2 \gamma  \sqrt{R_{c}} R_{\Lambda }}{\left(2 R_{c}+R_{\Lambda }+1\right) \left(\gamma 
	\sqrt{R_{c}}+\sqrt{R_{c}+R_{\Lambda }+1}\right)}+3(1+w)\bigg) &= 0 \label{rcl1} \\
-\left(2 \sqrt{R_{c}} \sqrt{R_{c}+R_{\Lambda }+1} \left(2 R_{\Lambda }-3 w-1\right)+\gamma(1
	 -R_{\Lambda })+3 \gamma  w \left(R_{\Lambda }+1\right)\right)&=\frac{2\gamma(3 w+1)}{3(w+1)} R_{c}R_{\Lambda} \label{rcl2}
\end{align}
\end{widetext}
These equations admit solutions where $R_{\Lambda}=0$, which always satisfies (\ref{rcl1}), leading to (\ref{rcl2}) taking the form 

\begin{align}
2 \gamma  R_{c}^2+\gamma  R_{c}+2 \sqrt{R_{c}+1} R_{c}^{3/2}+3 \sqrt{R_{c}+1}
\sqrt{R_{c}}-\gamma =0 , 
\end{align}
which possesses solutions $0 < R_{c} \sim \gamma^{2}/9 \ll 1$ for  $0<\gamma \ll 1$ (see Figure \ref{omegagamma}). We do not consider $\gamma <0$ here as it may be shown that solutions for $R_{c}$ are either negative or greater than or equal to $1/2$.

\begin{figure}
	\center
	\epsfig{file=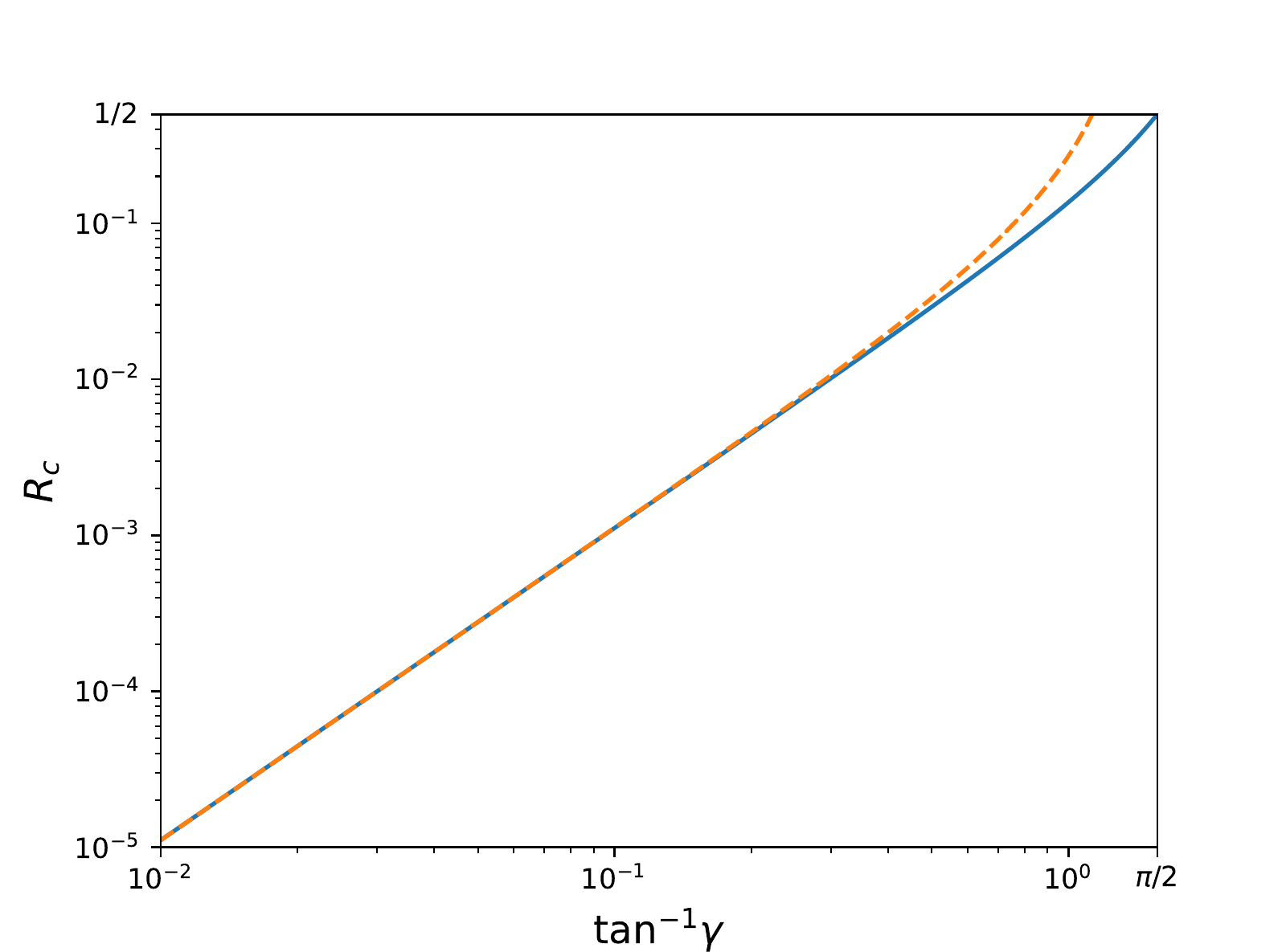,width=9.cm}
	\caption{Plot of a solution for $R_{c}=\Omega_{c}/\Omega_{w}$ as a function of $\tan^{-1}\gamma$, with $\Omega_{\Lambda}=0$. The solid curve represents that exact solution, which asymptotes to $R_{c}=1/2$ as $\gamma \rightarrow \infty$ and the dashed curve is the approximate solution $R_{c}= \gamma^{2}/9$, applicable for $\gamma \ll 1$. Interestingly, this soluton is independent of the equation of state $w$ of the dominant matter component.}
	\label{omegagamma}
\end{figure}

\subsection{Non-tracking solutions in the presence of more realistic matter content}\label{nontracksolutions}
For completeness, we now look at solutions to the field equations in the presence of realistic matter content (combined radiation and dust in typical abundances). The evolution of $c^{2}$ for several cosmological models is shown in Figure \ref{cosmoevolution}.

\begin{figure}
	\center
	\epsfig{file=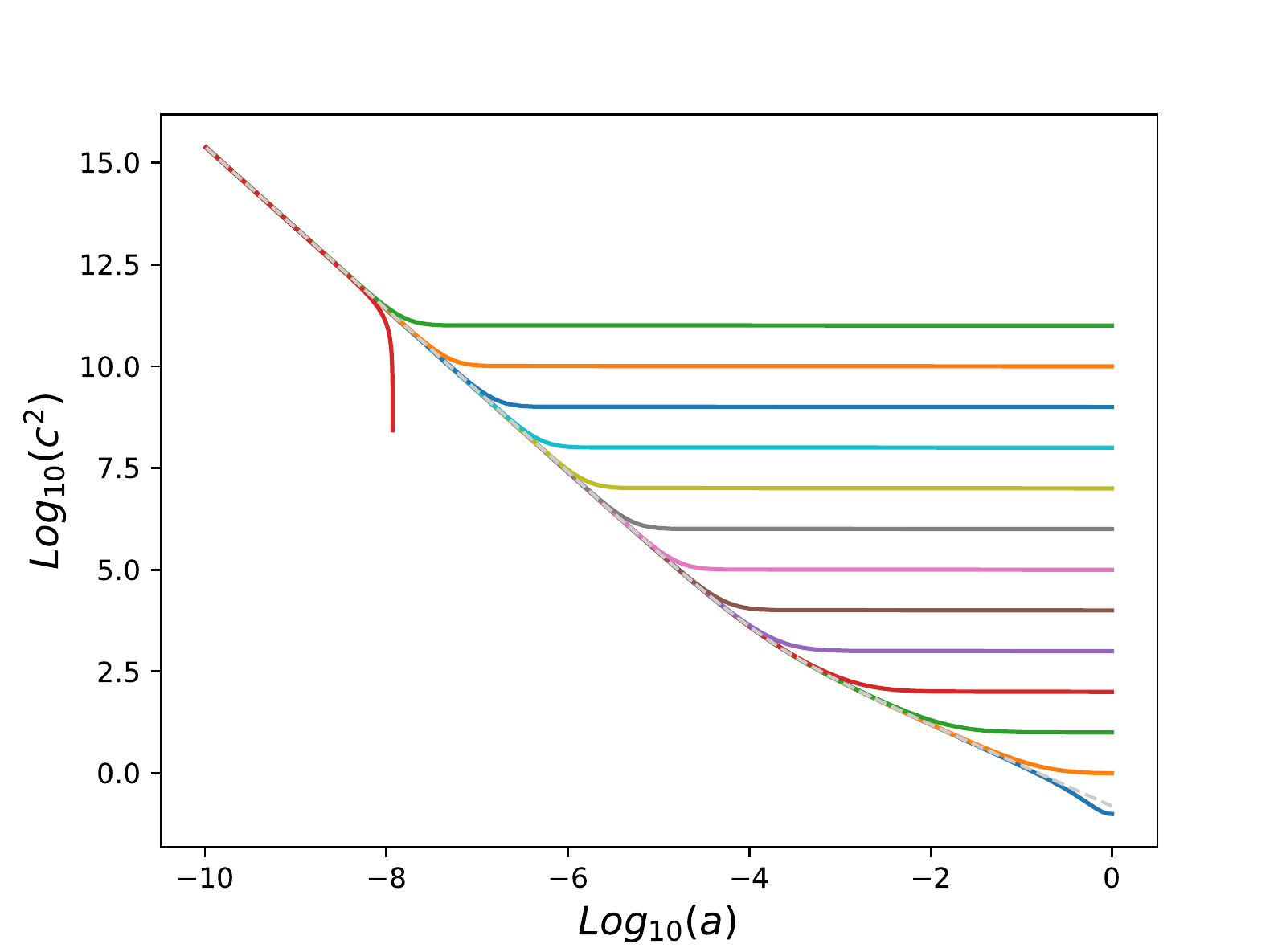,width=9.cm}
	\caption{Typical cosmological evolution of $c^{2}$ in the presence of radiation and dust. The dashed line is the tracking solution (\ref{infinitetrack}).}
	\label{cosmoevolution}
\end{figure}

We see that if $c^{2}$ begins above the tracking solution, it can stay close to the tracking solution for a time, eventually departing to take on a constant value. When this happens, the $c$ field - in having an effective density which scales as $c^{2}/a^{2}$ - gravitates in the same way as curvature. The departure from the tracking solution indicates that this effective curvature has begun to dominate the matter components (with the fields $c$ and $\Lambda$ then being well approximated by the results of Section \ref{solutions_without_matter}), leading to conflict with data if this happens too early. Closeness to the tracking solution yields in the radiation era an additional radiation component with half of the energy density of the combined photon and neutrino fluid which is also in conflict with data. 

We have therefore identified the culprit for the phenomenological troubles of these models. But we have also found how to fix the
problem. 
The existence at finite $\gamma$ of tracking solutions for which the energy density due to $c$ is much smaller than that due to radiation (see Fig.~\ref{omegagamma}) is ultimately what enables such models to result in a realistic cosmology. These will be explored in future work. It is conceivable that the new gravitational degrees of freedom are capable of producing the late-time acceleration of the universe. We emphasize that Fig.~\ref{omegagamma} and Fig.~\ref{cosmoevolution} display the evolution of parity-violating torsionful quantities and it is not clear whether counterparts could exist in a torsion-free scalar tensor theory.

Finally, we note that if $c^{2}$ begins {\it below} the tracking solution, it tends to result in Universes that do not persist to the present moment. To understand this, we note that the $c$ equation of motion can be written as follows:
\begin{align}
\frac{d c^{2}}{da} &=  -a^{3}\frac{\bigg(\Lambda (3p-\rho)+3\rho(\rho+3p)\bigg)}{\bigg(6b^{2}-3a^{2}\rho-a^{2}\Lambda\bigg)}
\end{align}
and so generally if $c^{2}$ has decreased to reach $0$, it will continue attempting to decrease which is an impossibility for $c \in \mathbb{R}$. Interestingly, similar behaviour was observed by Toloza and Zanelli \cite{Toloza:2013wi} for the analogue of the field $c$ in a theory which differs from the one under consideration here by fixing the $\Lambda$ in (\ref{palatini}) to be a constant \footnote{Similar `catastrophic complexification' of fields has also been observed in other modified theories of gravity \cite{Zlosnik:2017xpr}.}.

\section{Hamiltonian structure of the two branches}\label{Hamiltonian}

The Hamiltonian analysis of the system with $\gamma\rightarrow \infty$ sheds light on the nature of the two branches $c =Pa =0$ and $c\neq 0$. 
A simple way to derive the  Hamiltonian can be found in two steps applied to the action (\ref{SFRW1}) derived in Appendix I,
which is nothing but (\ref{SDcond}) specialized to FRW symmetry, with slightly different conventions and allowing for spatial curvature for generality.  

{\bf Step 1} We start by noting that the Euler term (i.e. the last two terms in (\ref{SFRW1})) can be rewritten in terms of the time derivative
of the Chern-Simons (CS) time of the system (see~\cite{alex0,LeeJ}). Indeed (\ref{SFRW1}) is equivalent to:
\bea
S^{g}\label{Faction1}
&=&\int dt \bigg(  2a^{2} \dot{b}+\frac{6}{\Lambda}\frac{d(bc^{2}-\frac{1}{3}b^{3}-bk)}{dt}\nn\\
&&
 + 2Na\bigg(k-c^2+b^2-\frac{\Lambda}{3}a^{2}\bigg)\bigg),
\eea
where indeed:
\be
\tau_{CS}= 6b(c^{2}-\frac{1}{3}b^{2}-k)
\ee
is the CS time of the system (i.e. the imaginary part of the CS functional~\cite{chopin,LeeJ}). We can therefore integrate the second term by parts to identify 
the momentum $\Pi$ conjugate to $\Lambda^{-1}$, and rephrase this in the form of a primary constraint 
forcing $\Pi$ to be proportional to the CS time seen as a function of connection variables $b$ and $c$. In other words, we 
can introduce a Lagrange multiplier $V$ to propose an action with the same 
equations of motion (since it is the same up to a boundary term):
\bea\label{Faction2}
S^{g'}
%\overset{b}{=} 
&=&\int dt \bigg(  2a^{2} \dot{b}+\Pi \frac{d\Lambda^{-1}}{dt}
 + 2Na\bigg(k-c^2+b^2-\frac{\Lambda}{3}a^{2}\bigg) \nn\\
&&+ V\bigg(\frac{\Pi}{6} +bc^{2}-bk-\frac{1}{3}b^{3}\bigg)\bigg).
\eea

{\bf Step 2} Having done this, we note that $c$ does not appear dotted or mutiplying any dotted quantity in the action, that is, 
it has no conjugate momentum. 
Within this approximation (spatial homogeneity and isotropy), this exposes the fact that $c$ is a connection degree of freedom 
which does not have a conjugate metric variable (there is no parity odd possible component for $e^a$). We could express this fact 
by introducing a term of the form $\dot c \Pi_c$ in the action and add and a further constraint forcing $\Pi_c\approx 0$ on-shell.
This would lead to a second-class constraint (as we shall see in Section~\ref{secondclass}).
Due to the complications inherent to second-class constraints we therefore opt to eliminate $c$ directly using the algebraic equation obtained by varying with respect to $c$:
\begin{align}
2Na c - bV c = 0 \label{ceq}
\end{align}

At once we see that this has two distinct branches: $c=0$ and $V = 2Na/b$ leading to two different theories.

{\bf Theory 1} If $c \neq 0$ then:
\be\label{VNeq}
V = N\frac{2a}{b}
\ee
and so the theory only has one independent constraint.  Eliminating $V$ we have:
\begin{align}
S^{g'}
&= \int dt \bigg(  2a^{2} \dot{b}+\Pi \frac{d\Lambda^{-1}}{dt}+ 2Na\bigg(\frac{2}{3}b^2-\frac{\Lambda}{3}a^{2}+\frac{\Pi}{6b}\bigg)\bigg) .
\end{align}
This is the theory containing the parity-odd solutions studied in this paper. Note that both the spatial curvature $k$ and $c$ disappear from the constraint. 

{\bf Theory 2} 
If $c=0$ then $V$ and $N$ remain independent:
\bea
S'
%&\overset{b}{=} 
&=&\int dt \bigg(  2a^{2} \dot{b}+\Pi \frac{d\Lambda^{-1}}{dt} + 2Na\bigg(k+b^2-\frac{\Lambda}{3}a^{2}\bigg) \nn\\
&&+ V\bigg(\frac{\Pi}{6} - bk-\frac{1}{3}b^{3}\bigg)\bigg) 
%\quad \mathrm{\textbf{Theory \quad 2}}
\eea
This is the theory studied in~\cite{alex2}, and it does not break parity.

We can write both Theory 1 and Theory 2 in the following form:
\begin{align}
S &=  \int dt \bigg( p \frac{d a^{2}}{dt}+\Pi \frac{d\Lambda^{-1}}{dt} - H (a^2,p;\Lambda^{-1},\Pi)\bigg)
\end{align}
where $p \equiv -2b$  and we have integrated by parts. Hence, the basic symplectic structure of both theories is:
\bea
\{a^2,p\}&=&1\label{PB1}\\
\{\Lambda^{-1},\Pi\}&=&1\label{PB2}
\eea
where the brackets denote the Poisson bracket. We now examine the structure of constraints of these two theories. 
We will distinguish between vacuum solutions and solutions where matter has been added in the form of either a dust fluid or a radiation fluid.
The details of the steps to the Hamiltonian formulation of these models are presented in Appendix II.

\subsection{Structure of Theory 2} 
For Theory 2 the Hamiltonian:
\be
H_{2} = -2N\sqrt{a^{2}}{\cal H}_{2} -V{\cal V}_{2}\\
\ee
is made up of two constraints:
\bea
{\cal H}_{2}&=&k+\bigg(\frac{p}{2} \bigg)^{2}-\frac{\Lambda}{3}a^{2} \label{hcon2}\\
{\cal V}_{2}&=&\frac{\Pi}{6}+\frac{p}{2} k  +\frac{1}{3}\bigg(\frac{p}{2}\bigg)^{3} \label{vcon2}
\eea
multiplied by appropriate Lagrange multipliers. 
The constraints form a closed algebra:
\begin{align}\label{com1}
\{{\cal V}_{2},{\cal H}_{2}\} &= \frac{\Lambda}{6}{\cal H}_{2} 
\end{align}
so they are functional first-class constraints (they close, but the ``structure constants'' are functions of phase space).
There is no need to add any further constraints to the total Hamiltonian to make the algebra close. 

The first constraint generates variations: 
\bea
\delta_N a^2 &=& \{a^{2},{\cal H}_{2}\} = \frac{p}{2} \nn\\
\delta_N p&=& \{p,{\cal H}_{2}\} =  \frac{\Lambda}{3}\nn\\
\delta_N \Lambda^{-1}&=&\{\Lambda^{-1},{\cal H}_{2}\} =  0\nn\\
\delta_N\Pi&=& \{\Pi,{\cal H}_{2}\} = \frac{\Lambda^{2}a^{2}}{3}
\eea
%[AND IT SHOULD BE POSSIBLE TO PROVE THAT THESE LEAD TO THE FIRST ORDER SYSTEM OF ODE DERIVED BEFORE,
%notice that $\delta_N$ is proportional to a time derivative]. 
The second constraint generates:
\begin{align}
\delta_V a^2= \{a^{2},{\cal V}_{2}\} &= \frac{k}{2} + \frac{1}{2}\left(\frac{p}{2}\right)^{2} \approx \frac{\Lambda}{6} a^{2}\nn\\
\delta_V p= \{p,{\cal V}_{2}\} &= 0 \nn\\
\delta_V \Lambda^{-1} = \{\Lambda^{-1},{\cal V}_{2}\} &= \frac{1}{6} \nn\\
\delta_V \Pi= \{\Pi,{\cal V}_{2}\} &=0.\label{confsmall}
\end{align}
The latter are shifts in $\Lambda$, together with a change in $a^2$ obtained from the Hamiltonian constraint. 
As we shall see in the next Section, the symmetry associated with these transformations is conformal symmetry. 
The first (Hamiltonian) transformations are a combination of a time lapse (resulting from the invariance under 
time reparameterizations) and a conformal transformation which would leave Lambda unchanged.

Thus the time evolution is given by the Hamiltonian evolution plus a conformal transformation. This 
is always the case with constrained systems (see~\cite{Dirac}; pp 11). Specifically, for a general quantity $f$
\be
\dot f=-2Na\{f,{\cal H}_2\}-V\{f,{\cal V}_2\}
\ee
so the time evolution equations with $N=1$ (proper time) are:
\bea
\dot  a&=&b-V\frac{\Lambda a}{3}\\
\dot b&=&\frac{\Lambda a}{3}\\
(\Lambda^{-1})^.=-\frac{\dot\Lambda}{\Lambda^2}&=&-\frac{V}{6}\\
\dot \Pi&=&\frac{2}{3}a^3\Lambda^2.
\eea
Together with ${\cal H}_2\approx 0$ and ${\cal V}_2\approx 0$, the full content of these is:
\bea
\dot a&=&b-\frac{\dot\Lambda}{2\Lambda}a\\
\dot b&=&\frac{\Lambda a}{3}
\eea
and they are  equivalent to the equations 31-34 of~\cite{alex2}. The first equation may also be written as 
$\dot a=b-Ta$ with $T=\dot \Lambda/(2\Lambda)$. 

Therefore, the arbitrariness of $\Lambda$ found in~\cite{alex2} merely reflects the fact that Lambda is pure gauge. 
The time evolution is given by the Hamiltonian evolution (keeping $\Lambda$ constant) plus a conformal transformation
resulting in whatever Lambda evolution we want. For a given $\Lambda(t)$ we should set
\be
V=6\frac{\dot\Lambda}{\Lambda^2}=12\frac{T}{\Lambda}.
\ee

\subsection{Structure of Theory 1}
For  Theory 1 we have:\begin{align}
H_{1} &=  -2N \sqrt{a^{2}}\bigg(\frac{2}{3}\bigg(\frac{p}{2}\bigg)^{2}-\frac{2\Pi}{6p}-\frac{\Lambda}{3}a^{2}\bigg)
\end{align}
with the single (Hamiltonian) constraint. The constraint generates transformations
\begin{align}
\delta_Na^2= \{a^{2},{\cal H}_{1}\} &= \frac{p}{3}+\frac{\Pi}{3p^{2}} \nn\\
\delta_N p = \{p,{\cal H}_{1}\} &=  \frac{\Lambda}{3}\nn\\
\delta_N \Lambda^{-1}= \{\Lambda^{-1},{\cal H}_{1}\} &= \frac{1}{3p} \nn\\
\delta_N\Pi=\{\Pi,{\cal H}_{1}\} &=  \frac{\Lambda^{2}a^{2}}{3}
\end{align}
and these result directly in the time evolution of the system. Choosing $N=1$ and evaluating 
$ \dot f=-2a\{f,{\cal H}_1\}$ for all quantities leads to the system of ODEs derived above.

\subsection{The structure of the theories in the presence of matter}
We now look at the effect of the inclusion of matter. For ease of illustration we will focus on the cases where the matter content is either in the form of a pressureless dust or a radiation fluid. This will be sufficient to make the point regarding the crucial difference between conformally invariant matter and non-conformally invariant matter. 

It can be seen using the results of Appendix II that the dust and radiation do not couple to $c$ and so equation (\ref{ceq}) still applies and leads to a similar phenomenon of branching depending on whether $c(t)=0$ or not and we will retain the designations \text{Theory 1} and \text{Theory 2} based on this distinction. We can write the combined matter and gravitational action as follows:

\begin{align}
S &=  \int dt \bigg( p \frac{d a^{2}}{dt}+\Pi \frac{d\Lambda^{-1}}{dt}+ {\cal P}\frac{d\phi}{dt} \nn\\
&- H (a^2,p;\Lambda^{-1},\Pi,\phi,{\cal P})\bigg)
\end{align}
where the symplectic structure is supplemented by the $\{\phi,{\cal P}\} = 1$ amongst matter field $\phi$ and its momentum ${\cal P}$.

\subsubsection{Theory 2 in the presence of dust matter} 
Using the results of Appendix II, it can be seen that for Theory 2 in the presence of dust we now have the Hamiltonian
\begin{align}
H_{2} &= -2N\sqrt{a^{2}}{\cal H}_{2}-V{\cal V}_{2}
\end{align}
where ${\cal V}_{2}$ retains the form (\ref{vcon2}) and now
\begin{align}
{\cal H}_{2} &= k + \bigg(\frac{p}{2}\bigg)^{2}-\frac{\Lambda}{3}a^{2}- \frac{{\cal P}}{2a}. 
\end{align}
Working out the Poisson bracket between the two constraints we now find:
\begin{align}
\{{\cal V}_{2},{\cal H}_{2}\} &= \frac{ \Lambda  \left(1 -\frac{3 {\cal P}}{ 4 a^{3}\Lambda }\right)}{6 }{\cal H}_{2}+\frac{\Lambda {\cal P} 
	\left(2 a^3 -\frac{3 {\cal P}}{\Lambda }\right)}{48 a^4}.
\end{align}
Therefore time evolution according to $H_{2}$ only preserves the constraints if, furthermore,  a secondary constraint

\begin{align}
{\cal W}_{2} \equiv 2 a^3 -\frac{3 {\cal P}}{\Lambda } = 0 \label{wcon2}
\end{align}
is additionally present. This constraint requires that $\Lambda$ be entirely fixed in terms of the dust momentum ${\cal P} =(2/3) \kappa \rho a^{3}$ which is equivalent to the result found in \cite{alex2}: one must have $\rho=\Lambda$ for $w=p/\rho=0$.
We further require that the constraint (\ref{wcon2}) is preserved by time evolution. It can be shown that $\{{\cal W}_{2},H_{2}\}\approx 0$ if
\begin{align}
V = -\frac{9p}{2a}\frac{N}{\Lambda}.
\end{align}
Therefore (as in Eq.~\ref{VNeq} for Theory 1) $V$ cannot be specified independently of $\{p,a,\Lambda,N\}$, signalling that the gauge freedom present in Theory 2 without matter is no longer present (as we will see the symmetry is conformal invariance, broken by dust).

There is, however, a function that has a weakly vanishing Poisson bracket with all the other constraints:
\begin{align}
{\cal Z}_{2} &= \frac{1}{Na}H_{2m}\bigg|_{V\rightarrow -(9p/2a)N/\Lambda} \\
&= -2{\cal H}_{2}+\frac{9}{2}\frac{p}{a^{2}\Lambda}{\cal V}_{2}.
\end{align}
On the constraint surface defined by the three constraints, we have that $\{{\cal Z}_{2},{\cal W}_{2}\} =\{{\cal Z}_{2},{\cal H}_{2}\} =\{{\cal Z}_{2},{\cal V}_{2}\} =0$.

Therefore, if we take constraints to be $({\cal V}_{2},{\cal W}_{2},{\cal Z}_{2})$ we form the algebra:
\begin{align}
\{{\cal Z}_{2},{\cal V}_{2}\}  &\approx 0 \\
\{{\cal Z}_{2},{\cal W}_{2}\}  &\approx 0 \\
\{{\cal V}_{2},{\cal W}_{2}\}  &\approx \frac{1}{108}a^{2}\Lambda^{3}
\end{align}
This suggests that in Theory 2 in the presence of pressureless dust there are two second-class constraints $({\cal V}_{2}, {\cal W}_{2})$ and one first-class constraint ${\cal Z}_{2}$.  We will see presently how this affects the counting of the number of degrees of freedom of the theory. 

\subsubsection{Theory 2 in the presence of a radiation fluid}
For Theory 2 in the presence of a radiation fluid we have the Hamiltonian

\begin{align}
H_{2} &= -2N\sqrt{a^{2}}{\cal H}_{2}-V{\cal V}_{2}
\end{align}
where ${\cal V}_{2}$ retains the form (\ref{vcon2}) and now

\begin{align}
{\cal H}_{2} &= k + \bigg(\frac{p}{2}\bigg)^{2}-\frac{\Lambda}{3}a^{2}-\chi \frac{{\cal P}^{4/3}}{a^{2}}
\end{align}
as explained in Appendix II. 
Working out the Poisson bracket between the two constraints we find:

\begin{align}
\{{\cal V}_{2},{\cal H}_{2}\} &= \frac{\Lambda}{6}{\cal H}_{2} + \frac{\Lambda}{6 a^{2}}\chi {\cal P}^{4/3}.
\end{align}
Therefore, time evolution according to $H_{2}$ only preserves the constraints if there is a secondary constraint:

\begin{align}
{\cal W}_{2}  \equiv {\cal P} =  0.
\end{align}
Recalling that $\rho\propto {\cal P}^{4/3}/a^4$ we see that this is nothing but the conclusion found in~\cite{alex2} that for $w=1/3$ 
the radiation density in the Universe is zero and thus the theory reduces to the vacuum theory (this follows directly from Eq.~(\ref{LeqF}), setting 
the right hand side to zero). 

Unsurprisingly we recover the structure of Theory 2 in vacuum. 
It can be checked that $\{{\cal W}_{2},{\cal V}_{2}\} = \{{\cal W}_{2},{\cal H}_{2}\} =0 $ and therefore  $V$ and $N$ remain arbitrary and independent functions of time. The Hamiltonian equation of motion for $\phi$ yields $\dot{\phi} \propto {\cal P}^{1/3} \approx 0$, showing that $\phi=\mathrm{cst.}$ on the constraint surface. Due to the global shift symmetry of the scalar field action of Appendix II, this constant can be set to zero without loss of generality, meaning that consistent dynamics takes place on a submanifold of phase space where ${\cal P} \approx 0$ and $\phi \approx 0$. Then, the symplectic structure $\{\phi,{\cal P}\}=1$ can be interpreted as non-commutation of constraints i.e. the presence of second-class constraints. We have $\{\phi,{\cal V}_{2}\}=0$ and $\{\phi,{\cal H}_{2}\}\approx 0$, therefore we have two first-class constraints $({\cal H}_{2},{\cal V}_{2})$ and two second-class constraints $(\phi,{\cal P})$.

\subsubsection{Theory 1 in the presence of matter}
It may be shown that in the presence of dust in the case of Theory 1 we now have:

\begin{align}
H^{d}_{1} &=  -2N \sqrt{a^{2}}\bigg(\frac{2}{3}\bigg(\frac{p}{2}\bigg)^{2}-\frac{2\Pi}{6p}-\frac{\Lambda}{3}a^{2}-\frac{\cal P}{2a}\bigg)
\end{align}
whereas in the presence of radiation we have

\begin{align}
H^{r}_{1} &=  -2N \sqrt{a^{2}}\bigg(\frac{2}{3}\bigg(\frac{p}{2}\bigg)^{2}-\frac{2\Pi}{6p}-\frac{\Lambda}{3}a^{2}-\chi\frac{{\cal P}^{4/3}}{a^{2}}\bigg)
\end{align}

As in the absence of matter, the theory contains one constraint and Hamilton's equations will be equivalent to equations (\ref{geneqi})-(\ref{geneqf}).
\subsection{Degrees of freedom}
Having found the constraint structure in the cases with and without matter (dust or radiation), we can now compute the degrees of freedom for each theory following the formula \cite{Dirac}:

\begin{align}
N_{dof} &= \frac{1}{2}\bigg(Dim_{ph}-2F-S\bigg)
\end{align}
where $Dim_{ph}$ is the dimensionality of the unconstrained phase space, $F$ is the number of first-class constraints, and $S$ is the number of second-class constraints. Without matter, phase space is coordinatized by $(a^{2},p,\Lambda^{-1},\Pi)$ and so $Dim_{ph}=4$. For Theory 2 we have two first-class constraints and no second-class constraints and hence $N_{dof} = 0$. For Theory 1 we have one first-class constraint and no second-class constraints and hence $N_{dof} = 1$. In the presence of either radiation or dust, the phase space can be coordinatized by $(a^{2},p,\Lambda^{-1},\Pi,\phi,{\cal P})$ and so $Dim_{ph}=6$. For Theory 2 in the presence of dust we have one first-class constraint and two second-class constraints and hence $N_{dof} = 1$ whilst for Theory 2 in the presence of radiation we have three two-class constraints, two second-class constraints and hence $N_{dof}=0$. For Theory 1 we have one first-class constraint and no second-class constraints and hence $N_{dof} = 2$.

\begin{table}
\begin{tabular}{|c || c | c | c | c |}
\hline
 &$Dim_{ph}  $ & $ F$ &$ S$  & $N_{dof}$ \\
\hline\hline 
 GR (no matter) & 2 &1  &0  &0  \\
\hline 
 GR (matter) & 4 & 1 & 0 &1  \\\hline 
 Theory 1 (no matter) & 4 & 1 & 0 & 1 \\\hline 
Theory 1 (w. matter) & 6 & 1 &0  & 2 \\\hline 
Theory 2 (no matter) & 4 & 2 & 0 & 0 \\\hline 
Theory 2 (w. dust) &  6& 1 & 2 & 1 \\\hline 
Theory 2 (w. radiation) & 6 & 2 & 2 & 0 \\\hline
\end{tabular}
\caption{The structure of phase space and its constraints for various theories, as discussed in this Section.
Within FRW symmetry, $Dim_{ph} $ is the dimensionality of the unconstrained phase space, $ F$  is the number of primary
constraints, $S$ the number of secondary constraints, and $N_{dof}$ the resulting number of degrees of freedom.\label{tab:table1}
 }
\end{table}

The situation can therefore be summarized in table~\ref{tab:table1}, where we have included standard GR for comparison. 
As we can see Theory 2 does not have any new degrees of freedom in addition to GR. In addition, if its matter content is radiation, 
the the theory is equivalent to GR without matter. On the contrary the parity violating Theory 1 does have a genuine new degree of freedom.
We will understand this better when we identify the origin of the new constraint in Section~\ref{confsym}. 

\subsection{Hamiltonian structure for general values of $\gamma$}
\label{finitegamham}
Even though we defer to future work a more complete study of finite $\gamma$, we add some comments here for completeness. 
For a general value of $\gamma$ the action (\ref{Faction2}) is modified in that the term multiplying $V$ in the Lagrangian now has additional terms in it and the equation of motion obtained from varying $c$ now takes the form

\begin{align}
V(b^{2}-c^{2}+k+2bc\gamma) &= 4ca \gamma N
\end{align}
Thus outside of the limit $|\gamma|\rightarrow \infty$ this equation is not automatically satisfied by the special solution $c(t)=0$ but would additionally require $b^{2}(t) = -k$ to leave $V$ undetermined, which from the $N$ equation of motion would then imply $\Lambda a^{2}$ being constrained to vanish. Outside of this special case, it is generally possible to solve for $c$ and $V$ from their combined equations of motion - as in Theory 1 in the $\gamma\rightarrow \infty$ limit - and yield a Hamiltonian formulation with one constraint.

\section{The conformal invariance of the theory}\label{confsym}
It was conjectured by Dirac~\cite{Dirac} that 
all primary \emph{and} all secondary first-class constraints generate gauge transformations\footnote{
A possible resolution to this issue may be found in~\cite{Castellani:1981us},  where an algorithm for creating all the generating functions associated with gauge symmetries is presented. In some exotic cases, secondary first-class constraints can end up not appearing in the gauge generators. Blagojevic in his textbook~\cite{Blagojevic} claims that for all known relevant physical applications, Dirac's conjecture holds.}.
%I think in our case we're ok with our DOF counting as we have no secondary first-class constraints by the end of the analysis].
Here we show that the new constraint ${\cal V}$ unveiled by the Hamiltonian analysis of the system is conformal invariance.

Under a conformal transformation:
\bea
\tilde e^a&=& \phi^{1/2} e^a\\
\tilde R^{ab}&=&R^{ab}\\
\tilde \Lambda&=&\frac{\Lambda}{\phi},
\eea
we have that the Lagrangian $L$ derived from the action (\ref{SDcond}), as in $S=\int L$, transforms up to a boundary term as: 
\be\label{Sconf}
\tilde L=\phi L.
\ee
%[If $\phi$ is a constant, otherwise it stays inside the $\int d^{4}x$]
The fact that the Lagrangian is not invariant (but only a conformal density) implies that the EOM obtained by integration by parts
(e.g. the $\omega^{ab}$ equation) will not generally be conformally invariant.  All the other equations, however, will be. 

Note that the torsion is not a conformal tensor, since:
\be
\tilde T^a= \phi^{1/2} \left(T^a +\frac{d\phi}{2 \phi}e^a\right),
\ee
however this does not spoil the conformal invariance (with weight 1) of the total action. Note also that in a FRW background it is
the parity-even component of the torsion that fails to be a conformal tensor:
\bea
\tilde T&=&\frac{1}{\sqrt{\phi}}\left( T -\frac{\dot\phi}{2\phi}\right)\label{Tconf}\\
\tilde P&=&\frac{P}{\sqrt{\phi}}.
\eea
Hence we see all the relevant quantities transform as:
\bea
d\tilde t&=&\sqrt{\phi}dt\\
\tilde a&=&\sqrt{\phi}a\\
\tilde b&=&b\\
\tilde c&=&c\\
\tilde\Lambda&=&\frac{\Lambda}{\phi},
\eea
(it is easy to see that $\tilde g=g/\sqrt\phi$ for the new Hubble parameter, $g$, defined in Eq.~(\ref{gfunc})). 
By setting $\phi=1+V\Lambda/6$, with $V\ll 1$, we see that these imply (\ref{confsmall}). 
%Then, from (\ref{} , we see that the FRW reduced action transforms as ***, as it should. The same happens to *** up to a surface term. 

%This is clearly the underlying symmetry behind the constraint ${\cal V}_2$ [AND NOW WE MUST EXPLAIN WHY IS IT THAT FOR THEORY 1 THIS IS ABSENT]. 

Having identified the symmetry behind the extra constraint of the system we may now investigate why this is preserved
when $c=0$, but broken otherwise. The reason is that, as already noted,  the gravitational Lagrangian is not strictly conformally invariant, but 
a density ($L^g$ is multiplied by $\phi$; cf. Eq.~(\ref{Sconf})). Thus, any field equations involving integrations by parts will not be conformally invariant. In our case the relevant equation is the connection equation (\ref{omEq}). Indeed it can be checked that in general
this is only conformally invariant for SD solutions (satisfying (\ref{SDequation})). For the FRW reduction, the
 connection equations are (\ref{TeqF})  and  (\ref{PeqF}). It can be 
seen that  equation (\ref{TeqF}) with no matter and $\gamma\rightarrow \infty$ becomes: 
\be
T=\frac{\dot \Lambda}{2\Lambda},
\ee
which 
is conformally invariant, even though its LHS and RHS are not. However equation (\ref{PeqF}) is not conformally invariant, unless $P=c=0$. If $c\neq 0$  this implies Eq.~(\ref{PeqF1}), obviously not conformally invariant. 
Hence the component of the connection equation that breaks conformal invariance is precisely the same that breaks parity.

This is why Theory 2 in the absence of matter is nothing but GR with one extra gauge degree of freedom represented by the cosmological constant
(see Table~\ref{tab:table1}). Discounting gauge degrees of freedom, the theory is the same as GR. In the same way that Minkowski and de Sitter space-times have no d.o.f., the theory does not have any. The apparent new degree of freedom is pure gauge, and this is why Lambda is 
left undetermined. The same happens if the matter content is conformally invariant (in which case the theory is equivalent to GR in vacuum, with the density of the matter forced to be zero for the equations of motion to admit consistent solutions). 
Addition of non-confomal matter adds a degree of freedom to both GR and Theory 2. 
In contrast, Theory 1 never displays conformal invariance at the level of the equations of motion. It always has one fewer constraint than Theory 2, and consequently one more degree of freedom than GR, with or without matter (conformally invariant or not).

Finally we remark that by choosing 
\be
\phi=\frac{\Lambda}{\Lambda_0}
\ee
it is possible to map theory into a constant $\Lambda$ theory with a varying gravitational constant, should the matter content be non-conformally invariant. 
This is indeed what is found in the solutions reported in~\cite{alex2}, for Theory 2, which does preserve conformal invariance.

\section{Parity violation as the source of a second-class constraint}\label{secondclass}
There is an alternative way to understand the double branch structure triggered by setting the parity violating component of the 
torsion $c$ to zero, or not. Here we will illustrate this ignoring matter (it is straightforward but tedious to include matter).

The fact that $c$ has a vanishing conjugate momentum (see Eq.~(\ref{Faction2})) could have been phrased as another constraint, rather
than solving for $c$ and eliminating it from the action, as we did in Section~\ref{Hamiltonian}, step 2. Then, we would have to 
contend with 3 constraints, encoded in action:
\bea\label{Faction3}
S^{g''}
%\overset{b}{=} 
&=&\int dt \bigg(  
p \frac{d a^{2}}{dt}+\Pi \frac{d\Lambda^{-1}}{dt}
+\dot c\Pi_c\nn\\
&&+2Na{\cal H}_3+V{\cal V}_3+\lambda\Pi_c
\bigg)
\eea
where
\bea
{\cal H}_3 &=& k-c^2+\frac{p^2}{4}-\frac{\Lambda}{3}a^{2} \\
{\cal V}_3&=& \frac{\Pi}{6}+\frac{p}{2} k  +\frac{1}{3}\bigg(\frac{p}{2}\bigg)^{3} 
-\frac{p}{2}c^2 
\eea
or in the Hamiltonian:
\be
H_3=-2Na{\cal H}_3-V{\cal V}_3-\lambda\Pi_c,
\ee
with $\{ c,\Pi_c\}=1$, beside the previously defined Poisson brackets, (\ref{PB1}) and (\ref{PB2}). 
Computing the algebra of constraints we find that, even without setting $c=0$, we still
have:
\be
\{{\cal V}_{3},{\cal H}_{3}\} = \frac{\Lambda}{6}{\cal H}_{3} ,
\ee
as in Theory 2 (see Eq.~\ref{com1}). In addition:
\bea
\{{\cal H}_3,\Pi_c\}&=&-2c\label{com31}\\
\{{\cal V}_3,\Pi_c\}&=&-pc.\label{com32}
\eea

There are now 2 ways to turn this system of second-class constraints into first-class constraints, resulting in theories 
with a different number of d.o.f. One is to {\it increase} the number of constraints by imposing the further constraint:
\be
c\approx 0,
\ee
so that now (\ref{com31}) and (\ref{com32}) vanish on-shell (or close within the system including the new constraint). 
Note that $\{c,{\cal H}_3\}=\{c,{\cal V}_3\}=0$, so the whole system is now first-class if we ignore $c$ and $\Pi_c$. Obviously, 
$c=0$ and  $\Pi_c=0$ do not commute, since $\{c,\Pi_c\}=1$, but this is simply two second-class constraints over two variables, 
rendering them irrelevant.  This results in Theory 2. 

The other way is to {\it decrease} the number of constraints and close  (\ref{com31}) and (\ref{com32}) by forming the linear combination of constraints ${\cal H}_3$ and ${\cal V}_3$ as they enter in the total Hamiltonian which does commute with $\Pi_c$.
This amounts to requesting that
\be
\{\Pi_c,-2Na{\cal H}_3-V{\cal V}_3\}=0
\ee
and from (\ref{com31}) and (\ref{com32}) this implies $N=-Vp/(4a)=Vb/(2a)$, so the linear combination sought is 
proportional to ${\cal H}_1$. This is Theory 1. 

We see that even though the counting leading to the number of degrees of freedom is done differently the final result is the same.
Instead of Table~\ref{tab:table1} (for no matter)  we now have Table~\ref{tab:table2}.

\begin{table}
\begin{tabular}{|l || c | c | c | c |}
\hline
 &$Dim_{ph}  $ & $ F$ &$ S$  & $N_{dof}$ \\
\hline\hline 
 Theory 1 (no matter) & 6 & 2 & 0 & 1 \\\hline 
%Theory 1 (w. matter) & 6 & 1 &0  & 2 \\\hline 
Theory 2 (no matter) & 6 & 2 & 2 & 0 \\\hline 
%Theory 2 (w. matter) &  6& 1 & 2 & 1 \\\hline 
\end{tabular}
\caption{The bifurcation associated with the $\Pi_c$ constraint, and how it results in the same number of degrees of freedom
for theories 1 and 2 (without matter) as in Table~\ref{tab:table1}. The situation is similar with the addition of matter.  \label{tab:table2}}
\end{table}

%What is the symmetry associate with this constraint? We see that it generates shifts in $c$, i.e. $\delta_\lambda c=
%\{c,\lambda \Pi_c\}= \lambda$. {\bf I suspect this is the symmetry behind the canonical transformation associated with the 
%Ashtekar formalism - SEE THE THIEMAN BOOK, THERE MIGHT BE SOMETHING THERE}. The symmetry is 
%\be
%A^i\rightarrow A^i + \gamma \epsilon ^{ijk}\omega^{jk}
%\ee
%or 
%\be
%\omega^{0i} \rightarrow  \omega^{0i} + \gamma \epsilon^{ijk}\omega^{jk}.
%\ee
%NO!

%\section{The finite $\gamma$ theory}
%[THERE IS A MYSTERY HERE. IN \cite{alex1} WE IGNORED POTRYAGIN BECAUSE IT IS ZERO FOR FRW WITH ZERO P.
%HOWEVER NOW WE DO NOT SEEM TO BE ABLE TO SET P TO ZERO IN THE FINITE GAMMA CASE, EXCEPT IN SOME
%TRIVIAL CASES ($\gamma+\rho=0$). WHAT IS GOING ON? WERE WE JUST STUPID?]

\section{Conclusions}
In this paper we found a new degree of freedom in homogeneous and isotropic, but parity violating Universes, allowing 
for Weyl curvature. It is present,
for example, in the theory proposed in~\cite{alex1}. At the level of homogeneous and isotropic space-times (for which GR has zero degrees of freedom without matter and one d.o.f. with matter) this is the only new degree of freedom of 
the theory proposed in~\cite{alex1}, but it would be interesting to study  the equivalent problem  for tensor and scalar perturbations, and see the implications
for the graviton. 

Within FRW models we found the new parity violating solutions, with and without matter, focusing on the theory without Pontryagin 
term ($\gamma\rightarrow\infty$) but with an eye on introducing it. The vacuum solutions reported in
Section~\ref{solutions_without_matter} are the non-SD solutions speculated to exist in~\cite{alex1,alex2}. In the presence of matter, we presented scaling solutions in Section~\ref{tracksolutions}, with disastrous phenomenology. 
The parity violating term enters the Hamiltonian constraint
with an effective energy density equal to one half that of the matter component. In Section~\ref{nontracksolutions} we found that in the presence of a mixture of matter and radiation
the field starts off by tracking radiation, but soon a cataclysm happens, with this term either 
behaving like a dominating curvature term, or leading to a complex manifold. 

However, this depends on the 
initial importance of the parity violating term. It turns out that  this depends  on the strength of the Pontryagin term, dialled by $\gamma$. 
As Fig.~\ref{omegagamma}  shows, the amount of parity violating scaling term increases with $\gamma$ and this is crucial. For infinite $\gamma$ it is 1/2 and this results in the double disaster depicted in Fig.~\ref{cosmoevolution}: either a curvature like domination or a complexification. But this can be averted with finite $\gamma$, leading to a realistic cosmology, including the 
late-time acceleration of the universe.  A full exploration of this very rich class of models will be presented in future work.

Having found these solutions in Sections~\ref{Hamiltonian}, \ref{confsym} and~\ref{secondclass} we sought to
to understand where the parity invariant solutions reported in~\cite{alex2} fit in with the parity breaking solutions reported here. 
By finding the Hamiltonian formulation of the theory in Section~\ref{Hamiltonian} we were able to determine that by setting $P=0$ or not, one is  led  to different theories, with different numbers of constraints and degrees of freedom. The counting is displayed
in Table~\ref{tab:table1}, where one sees that allowing for parity violating torsion results in a less constrained theory. 
In Section~\ref{confsym} we were able to identify the gauge symmetry associated with the extra constraint as conformal invariance. 
Switching on the parity breaking term in the torsion amounts to breaking conformal invariance at the level of the equations of motion, 
since the action is a conformal density, and the torsion equation involves an integration by parts, as explained in Section~\ref{confsym}. The solutions found in~\cite{alex2}, therefore, are nothing but conformal gauge transformations performed upon GR, unless non-conformal matter is added. In contrast the solutions reported in this paper represent a genuinely new degree of freedom, since conformal invariance is always broken.

In Section~\ref{secondclass} we present some final results on this bifurcation of a nominal single theory into two. 
Regardless of whether the parity violating term $P=ca$ is set to zero or not, it is generally true that its conjugate momentum,
$\Pi_c$, is zero. 
This is ultimately because there is no parity violating degree of freedom in the tetrad that can serve as conjugate variable
to the connection component associated with torsion $P$.
In view of this, we can either solve for $c$ and
find the double branch structure described in Section~\ref{Hamiltonian}; or else, instead regard $\Pi_c=0$ as a secondary constraint. The resulting
algebra can then be closed in two ways. If a further constraint is added, setting $P=ca=0$, then, together with $\Pi_c=0$, we end up with two second-class constraints (since $c=0$ and $\Pi_c=0$ do not commute) which commute with all the others. 
This is the hallmark of redundant variables, and renders
parity violation irrelevant to the theory, preserving conformal invariance. This is the theory of~\cite{alex2}. Alternatively we may close the algebra by linking the Hamitonian and conformal constraints, thereby breaking conformal invariance, but allowing for the parity violation term in the torsion. This results in the theory studied in this paper. It would be interesting to investigate the underlying symmetry corresponding to $\Pi_c=0$, if any, and how it interacts with conformal invariance. 

%Setting $c$ and $\Pi_c$ to zero just shows the consistency of theory 2. It makes the variables irrelevant via two second-class constraints which do not change the other constraints of the theory (Hamiltonian and conformal). 

With the benefit of hindsight it is easy to understand our results from considerations of symmetry and tensor algebra. 
After splitting space-time into time and space we are left with only two Cartesian tensors with which to build homogeneous and isotropic spatial structures: the Kronecker delta and the Levi-Civita tensors. In the second order formalism, with only the metric to deal with, there is no room for parity violation. This is because of the tensorial structure of the tetrad $e^a$: a set of one-forms with a single tetrad index. There is no isotropic way to link $e^i$ and $dx^j$ beside using $\delta^i_j$. Likewise for $e^0$ and $dt$ and $dx^i$. 

The first order formalism, and the possibility of torsion, opens up the doors to homogeneous, isotropic, but parity violating 
Universes. One can see this both from the spin-connection or the torsion viewpoints. The torsion is a set of 2-forms with a single 
tetrad index, $T^a$. This results in spatial $T^i$ which can be isotropically linked to $e^i e^j$ using $\epsilon^{ijk}$, 
beside the usual isotropic parity-even term $e^0 e^i$. Likewise, we have this parity-odd degree of freedom for a set of one-forms with two indices, such as the connection $\omega^{ab}$. From the definition of torsion we see that this degree of freedom in both quantities is the same. This degree of freedom propagates to the curvature (a set of 2-forms dependent on 2 tetrad indices) in the form of the Weyl tensor components (\ref{Weyl1}) and (\ref{Weyl2}). It is absent in the metric, and also in the matter content, since the stress-energy $\tau^a$ forms a set of 3-forms with a single tetrad index. An isotropic $\tau^0$ can only be proportional to 
$\epsilon_{ijk}e^i e^j e^k$ (the proportionality constant being the energy density); an isotropic $\tau^i$ can only be proportional 
to $\epsilon^{i}_{\ph{i}jk}e^0 e^j e^k$ (with pressure as the time-dependent proportionality constant).

We are therefore talking about a connection degree of freedom which does not have metric conjugate or a matter source. 
No wonder a different Hamiltonian structure is found, depending on whether we switch it on or off. 
Torsion and connection are the degrees of freedom that allow these parity violating models, and the results here add to existing examples of torsional degrees of freedom contributing towards the evolution of the Universe \cite{Mercuri:2009zt,Poplawski:2011j,Schucker:2011tc,Cid:2017wtf,Zlosnik:2018qvg}. They imply homogeneous and isotropic Weyl curvature. 
It would be interesting to explore, purely on the grounds of symmetry, what observables could detect this background effect (i.e. an effect present even before
adding perturbations), for example in the CMB or in weak lensing. The TB component of the polarization~\cite{TB,TB1}  and 
the vector part of the weak lensing~\cite{vector}  seem promising avenues to explore.

\section*{Acknowledgements}
We thank C. Contaldi, F. Hehl,  P. Jirou\v{s}ek, C. de Rham, A. Tolley, H. Westman, and T. Wiseman for helpful comments. TZ is funded by the European Research Council under the European Union's Seventh Framework Programme (FP7/2007-2013) / ERC Grant Agreement n. 617656 “Theories and Models of the Dark Sector:
DM, Dark Energy and Gravity”. JM was funded by the STFC Consolidated
Grant ST/L00044X/1.

\section*{Appendix I: FRW reduction of the action and parity transformations}
\label{frwactions}
In this appendix we provide a separate derivation of the field equations for spacetimes FRW symmetry. We first reduce the action to the form it takes in this symmetry and only then do variations and find the EOM. Obviously these are equivalent to those obtained from finding first the general EOM and then specialising  to FRW. However this approach produces results more easily adapted to  numerical work, as well as revealing the peculiar Hamiltonian structure of the system. 

Let us write  the FRW-reduced spin connection as
\begin{align}
\omega^{0i}&=bE^i \\
 \omega^{12}&=-\frac{K(r)}{r}E^2-cE^3 \\
\omega^{13}&=-\frac{K(r)}{r}E^3+cE^2\\ \omega^{23}&=-\frac{\cot\theta}{r}E^3-cE^1
\end{align}
and the co-tetrad $e^0=NE^0$ and $e^i=aE^i$ with the comoving basis one-forms:
\begin{align}
E^0=dt\quad E^1=\frac{dr}{K(r)}\quad E^2=rd\theta\quad E^3=r\sin\theta d\varphi.
\end{align}
Here $N$ is the lapse function and we have allowed for non-spatially flat FRW models ($K(r)=1/\sqrt{1-kr^2}$ with $k$ being the constant of spatial curvature). 
Then, the torsion is $T^{a} \equiv de^{a}+\omega^{a}_{\ph{a}b}e^{b}$ is given by:
\begin{align}
T^0 &=0 \\
T^i&=(\dot a-Nb)E^0E^i+ca\epsilon^i_{\ph ijk}E^jE^k \label{Apptorsion}
\end{align}
and the curvature two-form $R^{ab} \equiv d\omega^{ab}+\omega^{a}_{\ph{a}c}\omega^{cb}$ becomes:
\begin{align}
R^{0i}&=\dot bE^0E^i+bc\epsilon^i_{\ph ijk}E^jE^k\\ R^{ij}&=(k-c^2+b^2)E^iE^j-\dot c \epsilon^{ij}_{\ph{ij}k}E^0E^k
\end{align}
As announced in the main text (and obvious by comparing (\ref{Apptorsion}) with (\ref{Ti}) and (\ref{gfunc})) one can bridge $P$ and $g$ variables (used in previous literature) with $c$ and $b$ ones (more natural in this approach) according to 

\begin{align}
P &\equiv \frac{c}{a} \\
g  &\equiv \frac{b}{a}
\end{align}
It is now a matter of tedious calculation to prove that the gravitational action (\ref{SDcond}) when specialised to FRW becomes:

\begin{align}
\label{SFRW1}
S^{g}
&\overset{b}{=} \int dt \bigg( \bigg(2(k-c^2+b^2)Na + 2\dot{b}a^{2}\bigg)-\frac{2\Lambda}{3} Na^{3}  \nn\\
&-\frac{6}{\Lambda}(k-c^2+b^2)\bigg(\dot{b}-\frac{1}{\gamma}\dot{c}\bigg)+\frac{12}{\Lambda}bc\bigg(\dot{c}+\frac{1}{\gamma}\dot{b}\bigg)\bigg)
\end{align}
where $\overset{b}{=}$ means equal to up to a boundary term and the trivial integral over spatial coordinates has been omitted for notational compactness. Variation of the action (\ref{SFRW1}) yields the following equations of motion, where we have additionally allowed for the presence of perfect fluids in the field equations:

\begin{widetext}
	\begin{align}
	\frac{da}{dt}&= b-\gamma c +\frac{6(1+\gamma^{2})bc^{2}}{6\gamma bc+a^{2}(\Lambda+\kappa\rho)} \label{geneqi} \\
	\frac{db}{dt} &=  \frac{1}{3}a\bigg(\Lambda-\frac{\kappa}{2}(\rho+3 p)\bigg) \\
	\frac{d\Lambda}{dt} &= \frac{2\gamma ac\Lambda^{2}}{6\gamma bc + a^{2}(\Lambda+\kappa\rho)}\\
	\frac{dc}{dt} &= \frac{6 a b c (\kappa  \rho -2 \Lambda +3 \kappa  p)-a^3 \gamma  \kappa  (\rho  (\kappa  \rho -\Lambda )+3 p
		(\kappa  \rho +\Lambda ))}{6 \left(a^2 (\kappa  \rho +\Lambda )+6 b c \gamma \right)} \label{geneqf}
	\end{align}
\end{widetext}
For a quantity $v^{a}$ in the vector representation of the Lorentz group, a parity transformation can be represented by a matrix $\Lambda^{a}_{\ph{a}b} = \mathrm{diag}(1,-1,-1,-1)$ where $v'^{a}= \Lambda^{a}_{\ph{a}b}v^{b}$. Given this, the comoving basis one-forms will transform like $E^{0} \rightarrow E^{0}$, $E^{i} \rightarrow -E^{i}$ and the connection $\omega^{ab}$ will transform as $\omega^{0i}\rightarrow -\omega^{0i}$, $\omega^{ij} \rightarrow \omega^{ij}$. Hence we have $a\rightarrow a$, $b\rightarrow b $, $c\rightarrow -c$. We note from (\ref{geneqi}-(\ref{geneqf})) that the field $c$ and constant $\gamma$ appear only via combinations $c^{2}$,$\gamma^{2}$, and $c\gamma$. Therefore the effect of a parity transformation is equivalent to leaving $(a,b,c)$ invariant and taking $\gamma\rightarrow -\gamma$. The fact that solutions to the field equations for finite $\gamma$ don't generally agree for $\gamma\rightarrow -\gamma$ is a manifestion of parity violation in these models.

%\be\label{SFRW1}
%S^{g} =  \int \eps_{ijk}E^{0}E^{i}E^{j}E^{k}\bigg( \bigg(2(k-c^2+b^2)Na + 2\dot{b}a^{2}\bigg)-\frac{2\Lambda}{3} Na^{3}  
%-\frac{6}{\Lambda}\dot{b}(k-c^2+b^2)+\frac{12}{\Lambda}bc\dot{c}\bigg) 
%\ee
%

\section*{Appendix II: Models for matter actions}
Writing the gravitational action in FRW symmetry is a useful starting point for exploring the Hamiltonian formulation of the theory in this symmetry. It is further useful to add models of matter components in this context in order to see the effect of matter on the Hamiltonian formulation. We will consider two models in detail: a pressureless dust and a scalar field model acting as a radiation fluid.

\subsection{Pressureless dust}
An action describing pressureless dust is \cite{Brown:1994py}:
\begin{align}
S^d &=- 2\kappa\int \rho \bigg(\partial^{\mu}\phi\partial_\mu \phi+1\bigg)\sqrt{-g} d^{4}x.\\
    &= \frac{-2\kappa}{4!}\int \rho  \bigg(\partial^{\mu}\phi\partial_\mu \phi+1\bigg) \eps_{abcd}e^{a}e^{b}e^{c}e^{d}
\end{align}
where $\kappa \equiv 8\pi G$. The non-standard normalization of the action is due to choosing the Palatini Lagrangian to be $\eps_{abcd}e^{a}e^{b}\big(R^{cd}-\frac{\Lambda}{6}e^{c}e^{d}\big)$.
In FRW symmetry we have (again omitting the trivial spatial integration):
\begin{align}
S^d[\rho,\phi] &=- \int dt \frac{\kappa\rho}{3} \bigg(-\frac{1}{N^{2}}\dot{\phi}^{2} +1\bigg) N a^{3}.
\end{align}
First we introduce a new field $v = (1/N)\dot{\phi}$ so that the action becomes:

\begin{align}
S^d[\rho,\phi,v,{\cal P}] &= \int dt\frac{\kappa\rho}{3} \bigg(v^{2} -1\bigg) N a^{3} + N{\cal P}\bigg(\frac{1}{N}\dot{\phi}-v\bigg)
\end{align}
where ${\cal P}$ is a Lagrange multiplier enforcing the constraint between $v$ and $(1/N)\dot{\phi}$. We can then eliminate $v$ from the action principle by using its own equations of motion to get:

\begin{align}
S^d[\rho,\phi,{\cal P}] &= \int dt \bigg( {\cal P}\dot{\phi} - N a^{3}\frac{\kappa\rho}{3} -N \frac{3{\cal P}^{2}}{4 a^{3}\kappa\rho}\bigg)
\end{align}
We can furthermore eliminate $\rho$ via its equation of motion possessing the solution $\kappa\rho = 3{\cal P}/2a^{3}$ to yield:

\begin{align}
S^d[\phi,{\cal P}] &=  \int dt \bigg({\cal P}\dot{\phi}- N{\cal P}\bigg).
\end{align}
Thus we see that ${\cal P}$ is the canonical momentum of the dust field $\phi$.
\subsection{Radiation fluid}
It is known \cite{ArmendarizPicon:2000ah} that the following scalar field action in FRW produces the effect of a perfect fluid with equation of state $w=1/3$:

\begin{align}
S^{r} &= \xi \int (-\partial^{\mu}\varphi\partial_{\mu}\varphi)^{2}\sqrt{-g}d^{4}x \label{sr}
\end{align}
where the constant $\xi$ can be chosen so that the energy momentum due to this field is correctly normalized in Einstein's equations. Note that the the action (\ref{sr}) is manifestly invariant under local transformations $g_{\mu\nu}\rightarrow \phi(x^{\mu}) g_{\mu\nu}$, $\varphi \rightarrow \varphi$. In FRW symmetry we have (again omitting the trivial spatial integration):

\begin{align}
S^{r} &=  \frac{\xi}{3!}\int dt \frac{a^{3}}{N^{3}}\dot{\varphi}^{4}
\end{align}
As in the case of pressureless dust we may introduce a field $v\equiv \dot{\varphi}/N$ so that the Lagrangian becomes:

\begin{align}
S^{r} &=  \frac{\xi}{3!}\int dt \bigg(N a^{3}v^{4} + N{\cal P}\bigg(\frac{1}{N}\dot{\varphi}-v\bigg)\bigg)
\end{align}
Eliminating a real solution for $v$ using its own equation of motion, we recover the action

\begin{align}
S^{r}[\varphi,{\cal P}] &= \int dt \bigg({\cal P}\dot{\varphi}-2Na\bigg(\chi\frac{{\cal P}^{4/3}}{a^{2}}\bigg)\bigg)
\end{align}
where $\chi \equiv (3/8)(3/2\xi)^{1/3}$. From the way ${\cal P}$  enters the Hamiltonian constraint we see that:
\be
\frac{\kappa}{3}\rho=\chi \frac{{\cal P}^{4/3}}{a^4}. 
\ee

\bibliographystyle{unsrtnat}
\bibliography{references}

\end{document}